\documentclass[12pt,a4paper]{article}
\usepackage[utf8]{inputenc}
\usepackage[T1]{fontenc}
\usepackage{amsmath}
\usepackage{bm}
\usepackage{amssymb}
\usepackage{graphicx}
\usepackage{caption}
\usepackage{subcaption}
\usepackage[title]{appendix}
\usepackage{hyperref}
\usepackage{cite}
\usepackage{color}
\numberwithin{equation}{section}
\numberwithin{figure}{section}
\usepackage{etoolbox}

\makeatletter
\newcommand*{\lodbib@citeorder}{}
\newcommand*{\lodbib@notcited}{}% catch entries that were not cited

\def\citation{%
	\forcsvlist{\citation@i}}

\def\citation@i#1{%
	\ifinlist{#1}{\lodbib@citeorder}
	{}
	{\listxadd{\lodbib@citeorder}{#1}}}

\let\ltxorig@lbibitem\@lbibitem
\let\ltxorig@bibitem\@bibitem

\def\@lbibitem[#1]#2#3{%
	\csdef{lodbib@savedlabel@#2}{#1}%
	\@bibitem{#2}{#3}}

\def\@bibitem#1#2{%
	\xifinlist{#1}{\lodbib@citeorder}
	{}
	{\listadd{\lodbib@notcited}{#1}}%
	\csdef{lodbib@savedentry@#1}{#2}}

\renewenvironment{thebibliography}[1]
{\settowidth\labelwidth{\@biblabel{#1}}}
{\def\@noitemerr
	{\@latex@warning{Empty `thebibliography' environment}}%
	\section*{\refname}%
	\@mkboth{\MakeUppercase\refname}{\MakeUppercase\refname}%
	\list{\@biblabel{\@arabic\c@enumiv}}%
	{\leftmargin\labelwidth
		\advance\leftmargin\labelsep
		\@openbib@code
		\usecounter{enumiv}%
		\let\p@enumiv\@empty
		\renewcommand\theenumiv{\@arabic\c@enumiv}}%
	\sloppy
	\clubpenalty4000
	\@clubpenalty \clubpenalty
	\widowpenalty4000%
	\sfcode`\.\@m
	\lodbib@biblistloop
	\endlist}

\def\lodbib@biblistloop{%
	\forlistloop{\lodbib@bibitem}{\lodbib@citeorder}%
	\ifdefvoid{\lodbib@notcited}
	{}
	{\forlistloop{\lodbib@bibitem}{\lodbib@notcited}}}

\def\lodbib@bibitem#1{%
	\ifcsundef{lodbib@savedlabel@#1}
	{\ltxorig@bibitem{#1}}
	{\ltxorig@lbibitem[\csuse{lodbib@savedlabel@#1}]{#1}}%
	\csuse{lodbib@savedentry@#1}}
\makeatother

\begin{document}
	\title{Anomalous conductivities in the holographic St\"uckelberg model}
	%    \author{Eugenio Meg\'{i}as\thanks{\email{emegias}{ugr}{es}{}} \;and Nishal Rai
		%    \thanks{\email{nishalrai10}{gmail}{com}{}} \\
		%    Department of Physics}
	
	\author{Nishal Rai$^{1,2}$ and Eugenio Meg\'{\i}as$^{1,3}$}
	
	\date{
		\begin{small}
			$^{1}$ Departamento de F{\'\i}sica At\'omica, Molecular y Nuclear, \\ Universidad de Granada, Avenida de Fuente Nueva s/n, E-18071 Granada, Spain \\
			$^{2}$ Department of Physics, SRM University Sikkim, Upper Tadong, Sikkim, India \\
			$^{3}$ Instituto Carlos I de F{\'\i}sica Te\'orica y Computacional, Universidad de Granada, E-18071 Granada, Spain  \\
		\end{small}
		\vspace{0.5cm}
		\today
	}

	\maketitle
	\begin{abstract}
		We have studied a massive U(1) gauge holographic model with pure gauge and mixed gauge-gravitational Chern-Simons terms. The full backreaction of the gauge field on the metric tensor has been considered in order to explore the vortical and energy transport sector. The background solution has been computed numerically. On this background, we have considered the fluctuation of the fields and evaluated the different correlators. We have found that all the correlators depend on the mass of the gauge field.  Correlators such as the current-current one, $\langle J_xJ_x\rangle$, which were completely absent in the massless case,  in the presence of a finite gauge boson mass start picking up some finite value even at zero chemical potential. Similarly, the energy-current correlator, $\langle T_{0x}J_x\rangle$, which was also absent in the massless theory, has now a non-vanishing value but for finite values of the chemical potential. 
	%Using Kubo formulae we have evaluated the chiral magnetic and chiral vortical conductivities and studied their behaviour with the variation of the mass of the gauge field. 
	Our findings for the chiral vortical conductivity, $\sigma_V$, and the chiral magnetic/vortical conductivity of energy current, $\sigma^\varepsilon_B = \sigma^\varepsilon_V$, are completely new results. In addition to this, we have found that these anomalous transport coefficients depend linearly both on the pure Chern-Simon coupling, $\kappa$, and on the mixed gauge-gravity Chern-Simon coupling, $\lambda$. One of the results that we would like to highlight is that it’s not just $\kappa$ that contributes  to the $\sigma_B$  but there is an additional contribution from $\lambda$ as well 
	%In general, we see that all the anomalous transport coefficient depends on both  pure Chern-Simon coupling and the mixed gauge-gravity Chern-Simon coupling
	 unlike the previous studies.
	\end{abstract}

	%\pagebreak
	
	\clearpage

	\section{Introduction}
	The AdS/CFT correspondence \cite{Maldacena:1997re,Witten:1998qj} has been one of the most prominent theoretical handles for studying systems which were very hard to tackle previously. It states that, in the low energy limit, the large-$N_c$, ${\mathcal N}$ = 4 super Yang-Mills field theory in four-dimensional space is equivalent to the type IIB string theory in $AdS_5 \times S^5$  space. It has been widely applied for the study of strongly coupled systems such as condensed matter systems, QCD and hydrodynamics. Our current objective is to study the hydrodynamical approach using this correspondence.
	
	Quantum chiral anomalies are very fascinating properties which arise in the context of relativistic field theories of chiral fermions beyond perturbation theory~\cite{book1,book2,book3}. Chiral anomalies have played a very crucial role in the formulation of relativistic hydrodynamics \cite{Landsteiner:2012kd}. Anomaly-induced transport mechanisms have appeared on many occasions since the 80's~\cite{Vilenkin}. The axial current was the main topic in \cite{Newman:2005as}, and AdS/CFT correspondence was first used to anomalous hydrodynamics in \cite{Newman:2005hd}. Recently a lot of attention is gained by the effect of quantum anomalies on the hydrodynamics of otherwise conserved currents. The chiral magnetic effect~\cite{Fukushima:2008xe} and the chiral vortical effect~\cite{Kharzeev:2007tn,Banerjee:2008th,Son:2009tf,Landsteiner:2011cp} are two of such effects. In the former, the axial anomaly induces a current parallel to the external magnetic field, while in the latter a current is generated due to the presence of a vortex in the charged relativistic fluid. It has been argued that these and other anomaly-induced effects may be produced in non-central heavy ion collisions at RHIC and LHC~\cite{STAR:2009wot}, inducing in particular an event-by-event parity violation. These effects can also lead to anomalous transport properties in some condensed matter systems, such as the Weyl semi-metals~\cite{Basar:2013iaa,Landsteiner:2013sja}.
	
	In the past few years, these anomalous effects has been implemented in holography giving a lot of insights. One of such works is \cite{Gynther:2010ed}, where they considered a holographic model with a pure Chern-Simon term, and they computed the chiral magnetic conductivity which exactly matches with the results of the weakly coupled system. This is due to the fact that the anomalous conductivities have non-renormalization properties so that they are independent of the coupling constant. Later on, this model was extended to incorporate the effect of the energy-momentum tensor related to the energy current as well, and the mixed gauge-gravitational Chern-Simon term was added in the gravitational action~\cite{Landsteiner:2011iq,Landsteiner:2012dm,Megias:2013joa}. In these references the gauge fields were considered to be massless.  
	
	In a similar line of work, the authors of \cite{Jimenez-Alba:2014iia} have studied the dependence of the anomalous transport properties with the mass of the gauge field which is introduced via the St\"uckelberg mechanism. In their case, they have considered the probe limit. As a consequence the sectors comprising of the correlators related to the energy-momentum tensor were not accessible, in particular: i) the chiral vortical conductivity, ii) the chiral vortical conductivity of energy current, and iii) the chiral magnetic conductivity of energy current. In a sense, this model only comprises a pure gauge Chern-Simon term. Our goal in the present work is to access those sectors and to study the chiral vortical effects as well. To this end, we have considered the full backreaction of the gauge field onto the metric tensor, and included in the action of the model a mixed gauge-gravitational Chern-Simons term. 
	
	The paper has been organized as follows. In Section~\ref{sec:Holography_U1}, we will discuss the model under consideration and get the full backreacted numerical solution for the background. Next, we will discuss in Section~\ref{sec:Kubo_Formalae} the Kubo formulae and their relation with the retarded Green's functions, i.e. the correlators.  Using the AdS/CFT dictionary we will define these correlators in terms of the boundary terms. In Section~\ref{sec:Results} we will start presenting our results; first, we will compare the results with the known results for the massless case~\cite{Landsteiner:2011iq}, and after that, we will present our main results regarding the behaviour of the two-point correlators including the mass term for the gauge boson. We will discuss in the same section the effect of the mixed gauge-gravitational Chern-Simons term in these correlators, and finally we will show how the gauge boson mass affects the anomalous conductivities, namely the chiral vortical conductivity, $\sigma_V$, the chiral vortical conductivity of energy current, $\sigma^\varepsilon_V$, the chiral magnetic conductivity, $\sigma_B$, and the chiral magnetic conductivity for energy current, $\sigma^\varepsilon_B$.  Finally, we end with a discussion in Section~\ref{sec:Discussion}.

	\section{Holographic massive U(1) gauge theory}
        \label{sec:Holography_U1}

	We consider a holographic model with a massive U(1) gauge boson that includes both a pure gauge and a mixed gauge-gravitational Chern-Simon term in the action~\cite{Landsteiner:2011iq,Jimenez-Alba:2014iia}. The action of the model is
	\begin{eqnarray}
		S &=& \dfrac{1}{16 \pi G}\int d^5x\sqrt{-g}
		\Big[
		R+2\Lambda-\dfrac{1}{4} F_{MN} F^{MN} \nonumber \\
		&&-\dfrac{m^2}{2} (A_M-\partial_M\theta) (A^M-\partial^M\theta) \nonumber \\
		&&+ \epsilon^{MNPQR} (A_M-\partial_N\theta)\left(\dfrac{\kappa}{3}F_{NP}F_{QR}+\lambda R^A\,{}_{BNP}R^B\,{}_{AQR} \right)\Big] \nonumber \\
		&&+ S_{\rm GH}+S_{\rm CSK}  \,,  
		\label{act}
	\end{eqnarray}
	where
	\begin{eqnarray}
		S_{\rm GH} &=& \dfrac{1}{8 \pi G}\int_\partial d^4x    \sqrt{h} K \,,\\
		S_{\rm CSK} &=& -\dfrac{1}{2 \pi G}\int_\partial d^4x    \sqrt{h} K\lambda n_M\epsilon^{MNPQR}A_NK_{PL}D_QK^L_R \,,
	\end{eqnarray}
	are the Gibbons-Hawking boundary term, and a boundary term induced by the mixed gauge-gravitational anomaly, respectively, which have been well discussed in \cite{Landsteiner:2011iq}. $\theta$ is a field which ensures gauge invariance (up to gauge anomalies), and thus the mass term enters in a consistent way. As it mentioned in~\cite{Klebanov:2002gr,Gursoy:2014ela,Casero:2007ae}, the St\"uckelberg term arises as the holographic realization of dynamical anomalies.
        A comparison of the consistent form of the anomaly for chiral fermions~\cite{book1} with the variation of the action under axial gauge transformations, allows to fix the anomaly coefficients to
        \begin{equation}
          \kappa  = \kappa_p \equiv -\frac{G}{2\pi} \,, \qquad \lambda = \lambda_p \equiv -\frac{G}{48\pi} \,.
          \label{ano-coe}
        \end{equation}
        See e.g. Ref.~\cite{Landsteiner:2011iq} for a discussion. In the following we will refer to the values of Eq.~(\ref{ano-coe}) as the physical values of the anomaly coefficients.
        
	The bulk equations of motion for the action of Eq.~(\ref{act}) turn out to be
\begin{eqnarray}
\hspace{-0.7cm} G_{MN}-\Lambda g_{MN} &=& \dfrac{1}{2} F_{ML}F_N{}^L-\dfrac{1}{8} g_{MN} F^2 + \dfrac{m^2}{2} B_M B_N  	-\dfrac{m^2}{4} g_{MN} B_P B^P  \nonumber  \\
\hspace{-0.7cm}	&&+ 2\lambda \epsilon_{LPQR(M} \triangledown_B \left( F^{PL} R^{B}{}_{N)}{}^{QR} \right)\,, \label{eqten}
		\\
\hspace{-0.7cm}\triangledown_NF^{NM} &=& -\epsilon^{MNPQR} \left( \kappa F_{NP}F_{QR}+\lambda R^A{}_{BNP}R^B{}_{AQR} \right) + m^2 B^M \,, \label{eqmax}
        \end{eqnarray}
where we have defined a new field $B_M \equiv A_M-\partial_M\theta$, so that in the following $\theta$ will not appear explicitly anywhere. We have used the notation $X_{(MN)} \equiv \frac{1}{2}(X_{MN} + X_{NM})$.
	
	The ansatz for the background metric is a black hole solution in Fefferman-Graham coordinates, which is given by~\cite{Megias:2017czr,Megias:2019djo}
	\begin{equation}
		ds^2=-\dfrac{\ell^2}{\rho} g_{\tau\tau}(\rho) d\tau^2+\dfrac{\ell^2 }{\rho} g_{xx}(\rho)d\vec{x}^2+\dfrac{\ell^2 }{4\rho^2} d\rho^2,
		\label{bckmet}
	\end{equation}
where the boundary lies at $\rho=0$ and the horizon at $\rho = \rho_h$, while $\ell$ is the radius of AdS. The horizon $\rho_h$ is chosen in such a way that $g_{\tau\tau}(\rho_h)=0$, and the temperature of the black hole turns out to be
	\begin{equation}
		T=\dfrac{1}{2\pi}\sqrt{2\rho_h g_{\tau\tau}^{\prime\prime}(\rho_h)} \,.
	\end{equation}
The asymptotic expansion ($\rho\rightarrow0$)  of the solution of Eq.~(\ref{eqmax}) shows that the gauge field near the boundary behaves as
	\begin{equation}
		B_M(\rho)=a_0 \rho ^{-\frac{\Delta}{2} }+    a_1 \rho ^{\frac{\Delta }{2}+1} + \cdots \,, \label{eq:B_asymptotic}
	\end{equation} 
	where $m^2 \ell^2=\Delta(\Delta+2)$, with $\Delta$ the anomalous dimension of the dual current~\cite{Jimenez-Alba:2014iia}. The first(second) term in Eq.~(\ref{eq:B_asymptotic}) corresponds to a non-normalizable(normalizable) mode.  The scaling dimension of the normalizable mode is  ($3+\Delta$), and this puts an upper bound on the value $\Delta=1$. For $\Delta>1$ the dual operators become irrelevant (in the IR), and so we will be working in the range of values of $\Delta$ below this bound. 
	
%	\subsection{Holographic renormalization}

	%In order to account for chiral vortical effects we will be considering the full backreaction of the gauge field onto the metric.

	\subsection{Numerical solution for the background}

\begin{figure}
		\centering
		\includegraphics[width=0.55\linewidth]{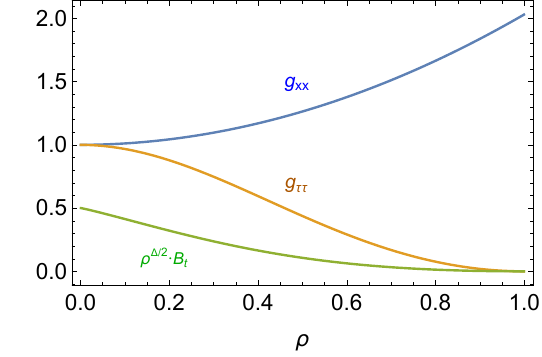}
		\caption{(color) Dependence of the background metric and gauge field with~$\rho$. We display the results for $g_{xx}(\rho)$ (blue), $g_{\tau\tau}(\rho)$ (orange),  and  $\rho^{\Delta/2} B_{t}(\rho)$ (green).  We have chosen $\Delta=0.1$ and $\mu_5=0.5$.}
		\label{fback}
	\end{figure}

	In order to account for the chiral vortical effects within the present model, we will be considering the full backreaction of the gauge field onto the metric. Plugging Eq.~(\ref{bckmet}) into Eqs.~(\ref{eqten}) and (\ref{eqmax}), the equations of motion for the background metric and gauge field turn out to be
	\begin{eqnarray}
		&&g_{xx}''(\rho )-\dfrac{g_{xx}'(\rho )}{\rho }    + \frac{1}{6\ell^2 \rho} \dfrac{ g_{xx}(\rho )}{g_{\tau \tau }(\rho )}\left(\dfrac{m^2 \ell^2
		}{4 } B_t(\rho )^2+\rho^2 B_t'(\rho )^2\right)
		=0 \,,  \label{eq:eom_gxx}\\
		&&g_{\tau \tau }'(\rho ) \left(1-\rho \dfrac{g_{xx}'(\rho)}{g_{xx}(\rho )} \right) + g_{\tau \tau }(\rho ) \dfrac{g_{xx}^\prime(\rho )}{g_{xx}(\rho )} \left(3 - \rho\dfrac{  g_{xx}^\prime(\rho
			)}{g_{xx}(\rho )}\right) \nonumber \\
		&& - \frac{1}{3} \dfrac{\rho ^2}{\ell^2} B_t^\prime(\rho)^2 + \dfrac{1}{12} m^2 B_t(\rho )^2=0 \,, \label{eq:eom_tt} \\
		&&B_t^{\prime\prime}(\rho ) + \frac{1}{2} \left( 3 \dfrac{g_{xx}^\prime(\rho )}{g_{xx}(\rho )}-\dfrac{g_{\tau\tau}^\prime(\rho )}{g_{\tau \tau }(\rho )}\right) B_t^\prime(\rho ) -\dfrac{\ell^2 m^2}{4 \rho^2}  B_t(\rho ) = 0 \,, \label{eq:eom_B}
	\end{eqnarray}
	where  the gauge field  has been chosen in the following way
	\begin{equation}
		B_M dx^M = B_t (\rho) dt \,,
	\end{equation}
so that $B_r=0$. We will solve numerically the above coupled differential equations with the following boundary conditions
\begin{equation}
 B_t(\rho_h)=0 \,, \hskip2cm \lim_{\rho \to 0} \left( \rho^{\Delta/2} B_{t}(\rho) \right) = \mu_5 \,, \label{eq:Bt_asymp}
\end{equation}
with $\mu_5$ being the source. As it is discussed in \cite{Jimenez-Alba:2014iia}, in the presence of a finite gauge boson mass $\mu_5$ does not correspond to a thermodynamic parameter, but it is instead a coupling in the Hamiltonian. As a result,  different values of chemical potential correspond to different theories. For completeness, we will provide the analytical solution of the background equations of motion~(\ref{eq:eom_gxx})-(\ref{eq:eom_B}) for vanishing $\mu_5$. These are
\begin{equation}
g_{\tau\tau}(\rho) = \frac{1}{\rho_h^2} \frac{(\rho_h^2 - \rho^2)^2}{\rho_h^2 + \rho^2} \,, \qquad g_{xx}(\rho) = 1 + \frac{\rho^2}{\rho_h^2}  \,, \qquad B_t(\rho) = 0 \,,  
\end{equation}
while the temperature turns out to be $T = \frac{1}{\pi}\sqrt{\frac{2}{\rho_h}}$. For the metric tensor, we demand that the solution is regular at the horizon, while at the boundary it reaches some constant value which we can always scale to set it to $1$.  Hereafter we will set the values of $\ell=1$ and $\rho_h=1$ for our numerical calculations, which one can fix as such by using the scaling symmetry of the metric tensor. This will set the units of all the quantities, i.e. $\mu_5$, $Q$, $M$, etc. We have plotted in Fig.~\ref{fback} the numerical solution of all the background fields, i.e. $g_{xx}(\rho)$, $g_{\tau\tau}(\rho)$ and $B_t(\rho)$.  One may note from this figure that $\lim_{\rho\rightarrow0}\left(\rho^{\Delta/2} B_{t}(\rho) \right) = \mu_5$.

	\section{Kubo formulae and correlators}
        \label{sec:Kubo_Formalae}

	In this section, we will discuss the Kubo formulae needed to compute the anomalous transport coefficients in our model, and set up the equations to evaluate these transport properties. The Kubo formulae for the anomalous conductivities have been well studied \cite{Amado:2011zx}. The authors of this reference have shown that the chiral vortical conductivity for charge and energy transport can be obtained from the following two-point functions
	\begin{equation}
		\begin{split}
			\sigma_V&=\lim_{k_c\rightarrow 0}\dfrac{i}{2 k_c}\sum_{a,b} \epsilon_{abc}\langle J^aT^{0b}\rangle\vert_{w=0}\,, \\
			\sigma_V^\varepsilon&=\lim_{k_c\rightarrow 0}\dfrac{i}{2 k_c}\sum_{a,b} \epsilon_{abc}\langle T^{0a}T^{0b}\rangle\vert_{w=0} \,,
		\end{split}
		\label{eqku1}
	\end{equation}
	where $\sigma_V$ is the chiral vortical conductivity and $\sigma^\varepsilon_V$ the chiral vortical conductivity of energy current, respectively. The chiral magnetic conductivities for charge, $\sigma_B$, and energy, $\sigma^\varepsilon_B$, current are given by
	\begin{equation}
		\begin{split}
			\sigma_B &= \lim_{k_c\rightarrow 0}\dfrac{i}{2 k_c}\sum_{a,b} \epsilon_{abc}\langle J^aJ^{b}\rangle\vert_{w=0} \,,\\
			\sigma_B^\varepsilon &= \lim_{k_c\rightarrow 0}\dfrac{i}{2 k_c}\sum_{a,b} \epsilon_{abc}\langle T^{0a}J^{b}\rangle\vert_{w=0}\,.
		\end{split}
		\label{eqku2}
	\end{equation}
	
	To compute these correlators one can use the AdS/CFT dictionary \cite{Landsteiner:2011iq,Son:2002sd,Herzog:2002pc}. Keeping this in mind, we proceed with the perturbation of the fields, where the background is set by the numerical solution as shown in Fig.~\ref{fback}. We will study the linear response of the fluctuation, so that we split the metric and gauge field into a background and a linear perturbation part, i.e.
	\begin{equation}
		g_{MN}=g_{MN}^{(0)}+\epsilon h_{MN}, \hskip2cm     B_{M}=B_{M}^{(0)}+\epsilon b_{M} \,.
		\label{eqpert}
	\end{equation}
	Then, we will follow the general procedure of Fourier mode decomposition \cite{Amado:2011zx}
	\begin{eqnarray}
		h_{MN}(\rho, x^\mu) &=& \int \dfrac{d^dk}{(2\pi)^d}h_{MN}(\rho)e^{-i\omega t+i \vec{k}.\vec{x}} \,,\\
		b_{M}(\rho, x^\mu) &=& \int \dfrac{d^dk}{(2\pi)^d}b_{M}(\rho)e^{-i\omega t+i \vec{k}.\vec{x}}\,.
	\end{eqnarray}
	Without the loss of generality, one can consider perturbations of frequency $\omega$ and momentum $k$ in the $z$-direction. In order to study the anomalous effect  we will switch on the fluctuations $B_i$, $h^i_{t}$ and  $h^i_{z}$, where $i= x,\; y$. Following this, we will substitute (\ref{eqpert}) in the equations of motion (\ref{eqten}) and (\ref{eqmax}), and consider the resulting expressions at order ${\mathcal O}(\epsilon)$.
	
	Since we are interested in computing correlators at zero frequency, we can set the frequency-dependent parts as zero in the equations, and solve the system up to first order in $k$. In this limit, the fields $h^i_z$ decouple from the system and take a constant value. Finally, we can write the system of differential equations for the shear sector as
\begin{eqnarray}
\hspace{-1.2cm} &&b_i^{\prime\prime}(\rho )+    \frac{1}{2} \left(\frac{g_{xx}^\prime(\rho )}{g_{xx}(\rho
				)}+\frac{g_{\tau \tau }^\prime(\rho )}{g_{\tau \tau }(\rho )}\right) b_i^\prime(\rho )  -\frac{\Delta  (\Delta
				+2)}{4 \rho ^2}  b_i(\rho )  \label{fleq1}\\
\hspace{-1.2cm}&&\qquad +\left(\frac{4 i \kappa  k \epsilon_{ij} b_j(\rho )}{\sqrt{g_{xx}(\rho ) g_{\tau \tau}(\rho )}}+\frac{g_{xx}(\rho )
				h^i{}_t^\prime(\rho )}{g_{\tau \tau }(\rho )}\right)B_t'(\rho )+  i \lambda k\epsilon_{ij}h^j{}_t'(\rho ) \Omega(\rho) = 0 \,,\nonumber\\
\hspace{-1.2cm}&&h^i{}_t^{\prime\prime}(\rho ) - \left(\dfrac{g_{\tau \tau }'(\rho )}{2 g_{\tau \tau }(\rho )} - \dfrac{5 g_{xx}'(\rho )}{2 g_{xx}(\rho)} + \dfrac{1}{\rho }\right) h^i{}_t^\prime(\rho )+\dfrac{\rho B_t'(\rho )}{g_{xx}(\rho )}  b_i^\prime(\rho ) \nonumber\\
\hspace{-1.2cm}&&\qquad+\dfrac{\Delta  (\Delta +2)  B_t(\rho )}{4 \rho  g_{xx}(\rho )} b_i(\rho ) + i \lambda k\epsilon_{ij} \Phi_j(\rho) =0 \,,\label{fleq2}
\end{eqnarray}
where $i, j = x, y$. The explicit expressions of the functions $\Omega(\rho)$ and $\Phi_j(\rho)$ are given in Appendix~\ref{App:A}. 
	
	Asymptotic analysis of the fluctuations near the boundary ($\rho \rightarrow 0$) up to the first subleading order shows
	\begin{eqnarray}
		b_i(\rho) &=& b_i^{(0)} \rho ^{-\frac{\Delta}{2} }+    b_i^{(1)} \rho ^{\frac{\Delta }{2}+1} + \cdots \,,\\
		h_{t}^i(\rho) &=& h_{t}^i{}^{(0)} +    h_{t}^i{}^{(1)} \rho^2 + \cdots  \,,
	\end{eqnarray}
where the leading order terms $b_i^{(0)}$ and $h_{t}^i{}^{(0)}$ are the sources.  From the holographic description of the correlation functions, one can evaluate the one-point functions as
\begin{eqnarray}
\langle J_a \rangle &=& \dfrac{\delta S_{\text{ren}}}{\delta b_a^{(0)}}=-\dfrac{2}{16\pi G}(\Delta+1)b_a^{(1)}\,, \qquad  (a = x,y) \,, \\
\langle T_{0a}\rangle&=&\dfrac{\delta S_{\text{ren}}}{\delta h_{t}^a{}^{(0)}}=\dfrac{1}{16\pi G}\left(2h_{t}^a{}^{(0)}+h_{t}^a{}^{(1)}\right) \,, \qquad (a=x,y) \,,
\end{eqnarray} 
where $S_{\textrm{ren}} = S + S_{\textrm{ct}}$ is the renormalized action, with $S$ the action given in Eq.~(\ref{act}) and $S_{\textrm{ct}}$ the counterterm. The procedure to evaluate this counterterm is given in \cite{Landsteiner:2011iq} and \cite{Jimenez-Alba:2014iia}. We find that the counterterm needed to renormalize this theory is the same as the one given in~\cite{Jimenez-Alba:2014iia}, i.e. the mixed gauge-gravitational Chern-Simons term does not introduce new divergences, and so the renormalization is not modified by it (see e.g. Ref.~\cite{Landsteiner:2011iq} for a discussion in the massless case).  In this regards, we are not writing the counterterm $S_{\textrm{ct}}$ explicitly.   $\langle J_a\rangle$ and $\langle T_{0a}\rangle$ correspond to current and energy-momentum tensor one-point functions, respectively~\footnote{$J_i$ and $ T_{0i}$ are related with the fluctuations $b_i$ and $h^i_t$, respectively, with $i= x, y$.}. Similarly, the two-point functions can be obtained by taking the variation of one-point function with respect to the corresponding source term, i.e.
	\begin{eqnarray}
		\langle J_aJ_b\rangle&=&  \dfrac{\delta \langle J_a\rangle }{\delta b_b^{(0)}} \,, \qquad (a, b = x,y) \,, \\
		%    \langle J_xJ_y\rangle&=& \dfrac{\delta \langle J_x\rangle }{\delta b_y^{(0)}}\,,  \qquad  \langle J_yJ_x\rangle = \dfrac{\delta \langle J_y\rangle }{\delta b_x^{(0)}}\,, \\
		\langle J_aT_{0b}\rangle&=& \dfrac{\delta \langle J_a\rangle}{\delta h^b_t{}^{(0)}} \,, \qquad (a, b = x,y) \,, \\
		\langle T_{0a} J_b\rangle&=& \dfrac{\delta \langle T_{0a}\rangle}{\delta b_b^{(0)}} \,, \qquad (a, b = x,y) \,, \\
		\langle T_{0a} T_{0b}\rangle&=& \dfrac{\delta \langle T_{0a}\rangle}{\delta h^b_t{}^{(0)}} \,, \qquad (a, b = x,y) \,. 
	\end{eqnarray}
	From the above expressions, it is clear that it is required the leading and subleading parts of the asymptotic expansion of the fluctuations to evaluate the two-point functions we are interested in. To do so we have solved numerically the coupled differential equations of the fluctuations (\ref{fleq1}) and (\ref{fleq2}) and imposed suitable boundary conditions, i.e. i) regularity at the horizon, and ii) sourceless condition at the asymptotic boundary.

\section{Results}
\label{sec:Results}

In this section, we will start presenting our results. Firstly, we will start with the massless case ($\Delta=0$) and compare the results with the previous work done in \cite{Landsteiner:2011iq}. In the second part of this section, we will consider the massive case $\Delta\ne 0$, and study the dependence of the two-point functions with $\Delta$ for different values of $\mu_5$. In both cases, we will set $G = 1/(16\pi)$ so that the physical values of the anomalous couplings are $\kappa=-1/(32\pi^2)$ and $\lambda=-1/(768\pi^2)$, cf. Eq.~(\ref{ano-coe}). Later on, we will study the dependence of the two-point functions with the parameters $\kappa$ and~$\lambda$. This is done to show that the parametric dependence of the correlators is linear in these parameters, but values of $\kappa$ and $\lambda$ different from $\kappa/\lambda=24$ are non-physical.  In addition to this, to make a direct comparison with the previous work in \cite{Jimenez-Alba:2014iia}, all the anomalous correlators have been displayed normalized by $|\kappa|^{-1}$.
	
		\begin{figure}[t]
		\centering
		\begin{subfigure}{0.45\textwidth}
			\centering
			\includegraphics[width=1\linewidth]{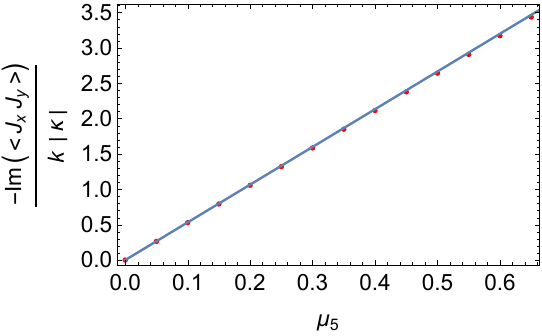}
		\end{subfigure}\hfill
		\begin{subfigure}{0.45\textwidth}
			\includegraphics[width=1\linewidth]{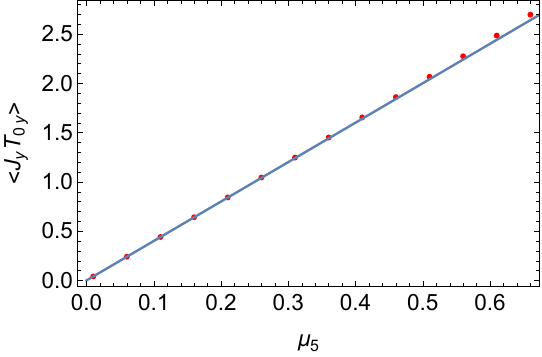}
		\end{subfigure}
		\begin{subfigure}{0.45\textwidth}
			\centering
			\includegraphics[width=1\linewidth]{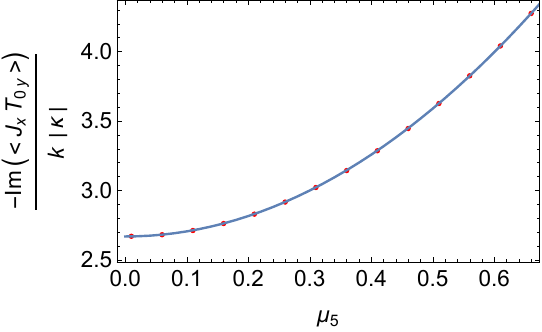}
		\end{subfigure}\hfill
		\begin{subfigure}{0.45\textwidth}
			\includegraphics[width=1\linewidth]{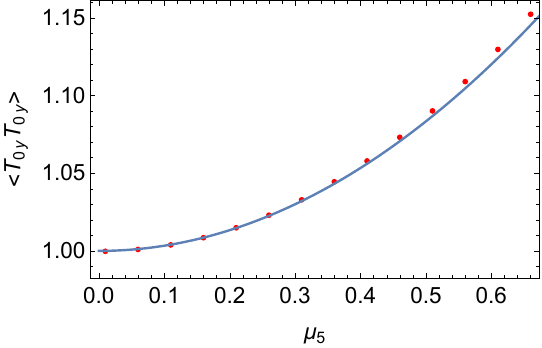}
		\end{subfigure}
		\caption{Upper panel: Plots of the correlators $\langle J_xJ_y\rangle $ (left) and $\langle J_yT_{0y}\rangle$ (right)  vs $\mu_5$. Lower panel: Plots of the correlators $\langle J_xT_{0y}\rangle $(left) and $\langle T_{0y}T_{0y}\rangle$ (right)  vs $\mu_5$. These plots are obtained in the massless case ($\Delta = 0$).}
		\label{figmu1}
	\end{figure}

	\subsection{Massless case}
        \label{subsec:massless}
        
\begin{figure}[t]
		\centering
		\includegraphics[width=.55\linewidth]{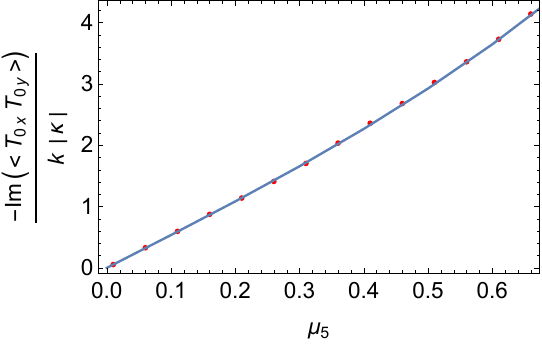}
		\caption{Plot of the correlator $\langle T_{0x}T_{0y}\rangle $ vs $\mu_5$ in the massless case $(\Delta = 0)$.}
		\label{figmu3}
	\end{figure}

	In the absence of mass, the correlators have been evaluated in \cite{Landsteiner:2011iq,Jimenez-Alba:2014iia}, leading to
	\begin{eqnarray}
		\langle J_xT_{0x} \rangle &=& \langle J_yT_{0y}\rangle =\dfrac{\sqrt{3}Q}{4\pi G \ell^3} \,, \nonumber \\
		\langle J_xJ_{y}\rangle &=& -\langle J_yJ_{x}\rangle = \kappa \dfrac{i  \sqrt{3}kQ}{2\pi G r_h^2} -\kappa \frac{i k \alpha}{6\pi G} = -i k \frac{\left(3 \mu_5 -\alpha \right)}{12\pi^2} \,, \nonumber\\
		\langle J_xT_{0y}\rangle &=& -\langle J_yT_{0x}\rangle=\langle T_{0x}J_y\rangle =-\langle T_{0y}J_x\rangle = \kappa \dfrac{3 i k Q^2}{4\pi G r_h^4} + \lambda\dfrac{2 i k \pi T^2}{ G } \,, \nonumber \\
                && \hspace{5.8cm} = -ik \left( \frac{\mu_5^2}{8\pi^2} + \frac{T^2}{24} \right) \,,  \label{cormu} \\
		\langle T_{0x}T_{0x}\rangle &=& \langle T_{0y} T_{0y}\rangle =\dfrac{M}{16\pi G \ell^3}\,, \nonumber  \\
		\langle T_{0x}T_{0y}\rangle &=& -\langle T_{0y}T_{0x}\rangle = \kappa \dfrac{i \sqrt{3}kQ^3}{2\pi G r_h^6} + \lambda \dfrac{4\pi i \sqrt{3}kQ T^2}{ G r_h^2} = -ik \left( \frac{\mu_5^3}{12\pi^2} + \frac{\mu_5 T^2}{12}  \right) \,,\nonumber
	\end{eqnarray}
	with $M=\dfrac{ r_h^4 }{\ell^2}+\dfrac{Q^2 }{ r_h^2}$ and $Q= \dfrac{\mu_5 r_h^2 }{\sqrt{3}}$ the mass and charge of the black hole solution computed in Poincar\'e coordinates, with blackening factor
        \begin{equation}
        f(r) = 1 - \frac{M \ell^2}{r^4} + \frac{Q^2 \ell^2}{r^6} \,.
        \end{equation}
        The Hawking temperature is given in terms of these black hole parameters as
        \begin{equation}
         T = \frac{r_h^2}{4\pi \ell^2} f^\prime(r_h) =  \frac{\left( 2 r_h^2 M - 3 Q^2 \right)}{2\pi r_h^5} \,.
        \end{equation}
The parameter $\alpha$ in Eq.~(\ref{cormu}) corresponds to the asymptotic value of the gauge field $A_t$ for $\rho \to 0$. In our case, we are assuming $\alpha = \mu_5$ for $\Delta = 0$, cf. Eq.~(\ref{eq:Bt_asymp}). The other correlators are vanishing in the massless case, i.e.
	\begin{equation}
		\langle J_x J_x \rangle = \langle J_y J_y \rangle = 0 \,, \qquad \langle T_{0x} J_x \rangle = \langle T_{0y} J_y\rangle = 0 \,. 
	\end{equation}
	While the correlators with the same indices are not induced by quantum anomalies (i.e. they are non-anomalous) and they become real, the correlators with different indexes are anomalous and they become imaginary. We will be comparing the numerical results with the analytical expressions given in the above equations, Eq.~(\ref{cormu}). We plot in Figs.~\ref{figmu1} and~\ref{figmu3} five independent non-vanishing correlators, while the other correlators are related to them through the expressions given in Eq.~(\ref{cormu}). In these and subsequent plots, it is understood that it has been taken the limit $k \to 0$ with $k \equiv k_z$. In these figures the dots stand for the numerical results, and the solid lines correspond to the analytic results of Eq.~(\ref{cormu}). One may observe that the numerical results are in good agreement  with the analytic expression.

	\subsection{Massive case}

	We will split our discussion into anomalous and non-anomalous correlators. We have found that the above mentioned relations between different correlators still hold in the massive case, i.e.  
	\begin{eqnarray}
		\begin{split}
		  \langle J_xT_{0x} \rangle &=\langle J_yT_{0y}\rangle \,, \\
                  \langle J_xJ_{y}\rangle &=-\langle J_yJ_{x}\rangle \,,\\
	           \langle J_xT_{0y}\rangle &=-\langle J_yT_{0x}\rangle=\langle T_{0x}J_y\rangle =-\langle T_{0y}J_x\rangle \,, \\        
		  \langle T_{0x}T_{0x}\rangle &=\langle T_{0y} T_{0y}\rangle \,, \\
	          \langle T_{0x}T_{0y}\rangle &= -\langle T_{0y}T_{0x}\rangle \,.
		\end{split}
	\end{eqnarray}
 In addition to this, there are two more independent correlators, i.e $\langle T_{0x}J_x \rangle=\langle T_{0y}J_y \rangle$ and $\langle J_{x}J_x \rangle=\langle J_{y}J_y \rangle$. In this regard, we will be plotting only seven independent correlators.
	
 \subsubsection{Non-anomalous correlators}
 \label{subsec:non_anomalous} 
	
	\begin{figure}[t]
		\centering
		\begin{subfigure}{0.45\textwidth}
			\centering
			\includegraphics[width=1\linewidth]{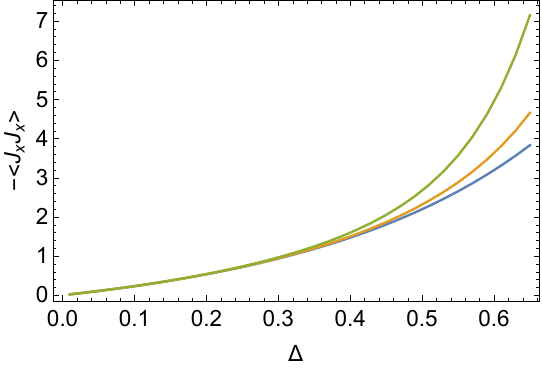}
		\end{subfigure}\hfill
		\begin{subfigure}{0.45\textwidth}
			\includegraphics[width=1\linewidth]{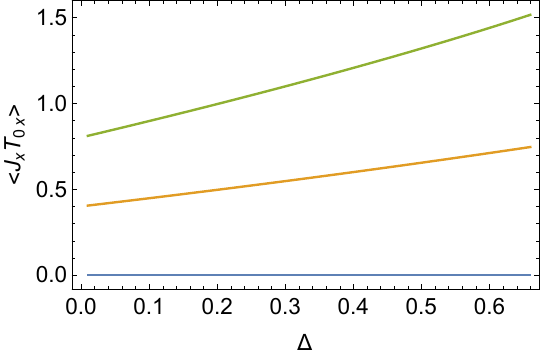}
		\end{subfigure}
		\centering
		\begin{subfigure}{0.45\textwidth}
			\centering
			\includegraphics[width=1\linewidth]{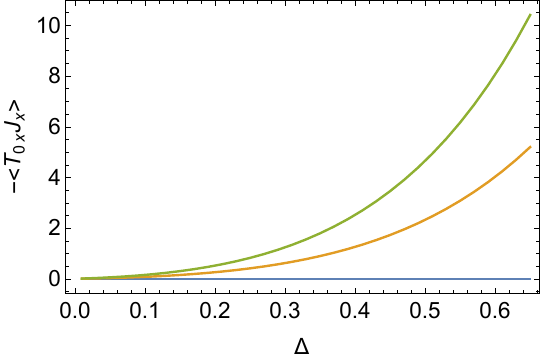}
		\end{subfigure}\hfill
		\begin{subfigure}{0.45\textwidth}
			\includegraphics[width=1\linewidth]{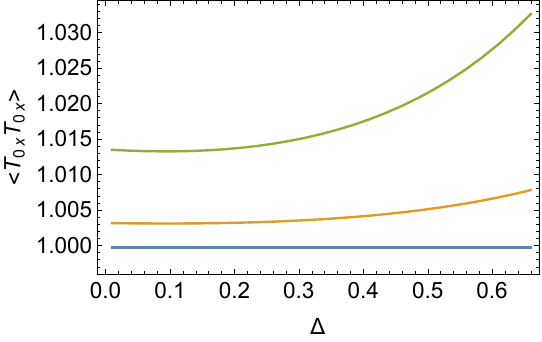}
		\end{subfigure}
		\caption{(color) Plots for non-anomalous correlators vs $\Delta$. Upper panel: Plot of the correlators $\langle J_xJ_{x}\rangle $ (left) and $\langle J_xT_{0x}\rangle$ (right)  vs $\Delta$. Lower panel: Plot of the correlators $\langle T_{0x}J_{x}\rangle $ (left) and $\langle T_{0x}T_{0x}\rangle$  (right) vs $\Delta$. We have considered in all the panels $\mu_5=\{0, 0.1, 0.2\}$ (blue, orange and green).}
		\label{figdl1}
	\end{figure}
	While the correlator $\langle J_xJ_x\rangle$ is vanishing for $\Delta=0$ (cf. Section~\ref{subsec:massless}), we can see from the Fig.~\ref{figdl1} (upper-left panel) that this correlator starts picking up some finite value in the massive case $(\Delta \ne 0)$.  With the increase of $\Delta$ the absolute value of this correlator increases quite sharply, and gets even shaper with the increase in $\mu_5$. This property, i.e. an increasing value of the (absolute value of the) correlator for increasing $\Delta$ and for finite $\mu_5$, is a general feature for all the non-anomalous coefficients as we will discuss below.
	
	%From the plot one can see that the value of 
	% We have made a quadratic fit of $\langle J_xJ_x\rangle$ as a function of $\Delta$ for a different value of $\mu_5$ which is given by

        We can see from  Fig.~\ref{figdl1} (upper-right panel) that for $\mu_5=0$ the correlator  $\langle J_xT_{0x}\rangle $ is zero for all values of $\Delta$.  As the value of $\mu_5$ increases, $\langle J_xT_{0x}\rangle$ becomes finite and its value increases with $\Delta$ in a somewhat linear fashion. The slope of $\langle J_xT_{0x}\rangle$ vs $\Delta$ also increases with the increase of $\mu_5$.
	
        In Fig.~\ref{figdl1} (lower panel-left) we can see that even though the correlator $\langle T_{0x}J_{x}\rangle $ is vanishing for $\Delta=0$, for finite values of $\Delta$ and $\mu_5$ this correlator is non-vanishing.  More in details, for a given finite value of $\mu_5$, the absolute value $|\langle T_{0x}J_{x}\rangle|$ increases quite sharply with $\Delta$. Notice that $\langle T_{0x}J_{x}\rangle$ was completely absent in the previous work \cite{Landsteiner:2011iq}, but we find now that it is non-vanishing at finite $\mu_5$ in the massive theory.
	
   Finally, we can see in Fig.~\ref{figdl1} (lower-right panel) that $\langle T_{0x}T_{0x}\rangle $ is independent of $\Delta$ for $\mu_5=0$, i.e. it has a constant value corresponding to the pressure term, a feature that has been well discussed in \cite{Gynther:2010ed,Landsteiner:2011iq,Landsteiner:2012dm, Jimenez-Alba:2014iia}. At finite chemical potential, this correlator increases with $\Delta$, a behavior which is sharper for larger values of $\mu_5$.

   \subsubsection{Anomalous correlators}
   \label{subsec:anomalous}

		\begin{figure}[t]
		\centering
		\begin{subfigure}{0.45\textwidth}
			\centering
			\includegraphics[width=1\linewidth]{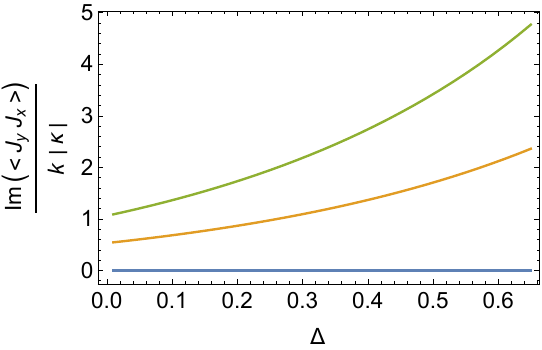}
		\end{subfigure}\hfill
		\begin{subfigure}{0.47\textwidth}
			\includegraphics[width=1\linewidth]{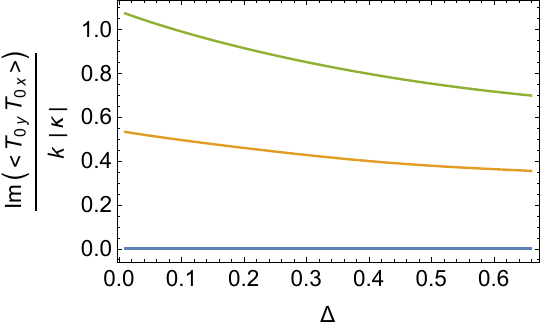}
		\end{subfigure}
		\caption{(color) Plot of the correlators $\langle J_{y}J_{x}\rangle $ (left) and $\langle T_{0y}T_{0x}\rangle $ (right) vs $\Delta$ with $\mu_5=\{0, 0.1, 0.2\}$ (blue, orange and green).}
		\label{figdl3}
	        \end{figure}
We display in Fig.~\ref{figdl3} (left) the behaviour of $\langle J_yJ_{x}\rangle $ vs $\Delta$. One can see that the absolute value of this correlator increases with $\Delta$, and the change is quite subtle. It is plotted in Fig.~\ref{figdl3} (right) the correlator $\langle T_{0y}T_{0x}\rangle$  vs $\Delta$, and unlike the other correlator, its absolute value decreases with the increase of $\Delta$.
		
	\begin{figure}
		\centering
		\includegraphics[width=0.5\linewidth]{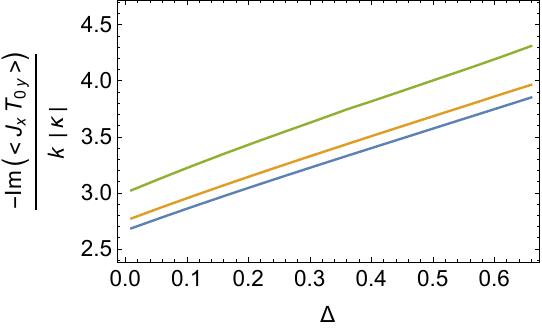}
		\caption{ (color) Plot of the correlator $\langle J_{x}T_{0y}\rangle $ vs $\Delta$ with $\mu_5=\{0, 0.15, 0.3\}$ (blue, orange and green).}
		\label{figdl4}
	\end{figure}
		In Fig.~\ref{figdl4} we have plotted  $\langle J_{x}T_{0y}\rangle $ vs $\Delta$. We find that the absolute value of this correlator increases with the increase in $\Delta$ and $\mu_5$. We have taken a different value of $\mu_5$ as compared to the other correlators, because for those values of $\mu_5$ the correlator did not have any substantial changes. The new values of $\mu_5=\{0, 0.15, 0.3\}$  are chosen to make these changes distinct in the figure.  We can see from the figure that even in the absence of $\mu_5$ this correlator is non-zero. This can be traced back to the temperature term, as the temperature does not vanish for $\mu_5=0$. 
	Finally, one may notice that in all the cases the values of two point correlators tend toward the analytic values as given in (\ref{cormu}) when considering the limit $\Delta \rightarrow 0$. This is also shown in the figures for the massless case.
	%%%%FIGURES FOR LAMBDA VARIATION START

	\begin{figure}[t]
		\centering
		\begin{subfigure}{0.45\textwidth}
			\centering
			\includegraphics[width=1\linewidth]{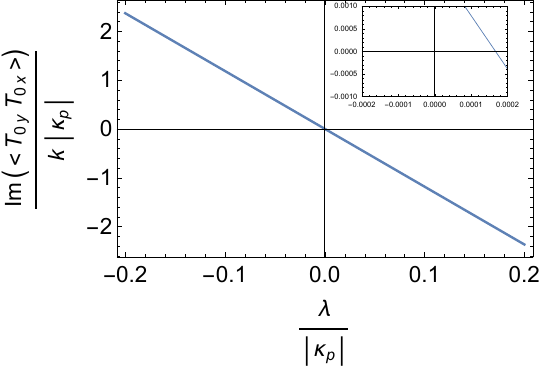}
		\end{subfigure}\hfill
		\begin{subfigure}{0.45\textwidth}
			\includegraphics[width=1\linewidth]{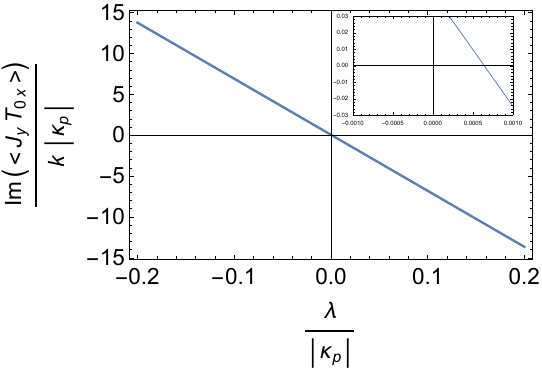}
		\end{subfigure}
		\caption{ Plot of the correlator $\langle T_{0y}T_{0x}\rangle $ (left) and $\langle J_{y}T_{0x}\rangle $ (right) vs $\lambda/|\kappa_p|$. The inset figures correspond to zooms of the main figures in the small $\lambda$ regime. We have considered in both panels, $\mu_5=0.1$,  $\Delta=0.1$ and $\kappa_p =-1/(32\pi^2)$.}
		\label{figld3}
	\end{figure}
	
	\begin{figure}[t]
		\centering
		\begin{subfigure}{0.45\textwidth}
			\centering
			\includegraphics[width=1\linewidth]{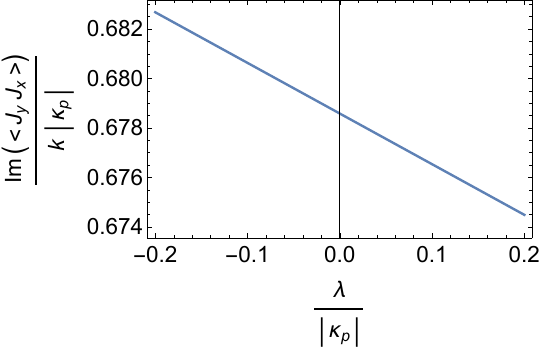}
		\end{subfigure}\hfill
		\begin{subfigure}{0.45\textwidth}
			\includegraphics[width=1\linewidth]{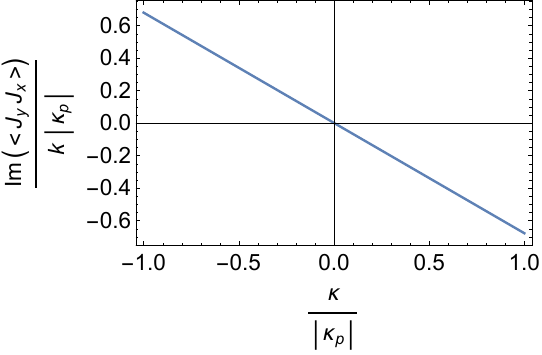}
		\end{subfigure}
		\caption{Left: Plot of the correlator $\langle J_{y}J_{x}\rangle $ vs $\lambda/|\kappa_p|$. Right: Plot of the correlator $\langle J_{y}J_{x}\rangle $ vs $\kappa/|\kappa_p|$ for $\lambda=-1/(768\pi^2)$.  We have considered in both panels $\mu_5=0.1$ and $\Delta=0.1$, while $\kappa_p=-1/(32\pi^2)$.} 
		\label{figld4}
	\end{figure}
	%%%%FIGURES FOR LAMBDA VARIATION END
	\begin{figure}[t]
		\centering
		\begin{subfigure}{0.45\textwidth}
			\centering
			\includegraphics[width=1\linewidth]{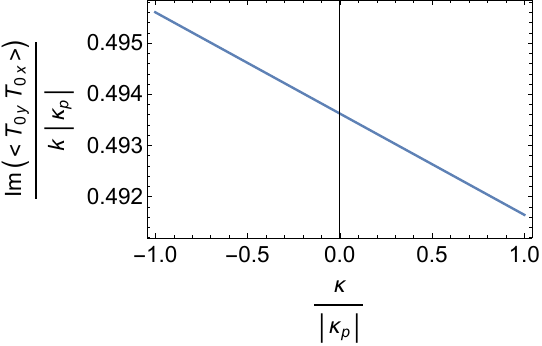}
		\end{subfigure}\hfill
		\begin{subfigure}{0.45\textwidth}
			\includegraphics[width=1\linewidth]{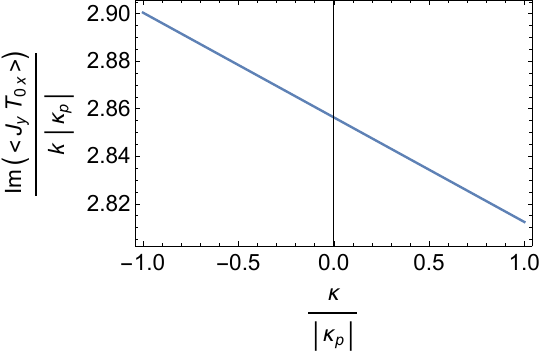}
		\end{subfigure}
		\caption{Plot of the correlator $\langle T_{0y}T_{0x}\rangle $ (left) and  $\langle J_{y}T_{0x}\rangle $ (right) vs $\kappa/|\kappa_p|$. In both cases, $\mu_5=0.1$, $\Delta=0.1$, |$\kappa_p|=1/(32\pi^2)$ and $\lambda=-1/(768\pi^2)$.}
		\label{figkp1}
	\end{figure}
	
	\paragraph{$\lambda$ and $\kappa$ dependence:}
	To study the dependence of the two-point functions with the parameter $\lambda$, we will consider the case where we fix the values as $\mu_5=\Delta=0.1$ and $\kappa=-1/(32\pi^2)$, and vary $\lambda$. Here we will only present the correlators that have a dependence on $\lambda$, while the $\lambda$ independent correlators are given in Fig.~\ref{figld2} of Appendix~\ref{App:B}. We have plotted in Figs.~\ref{figld3} and~\ref{figld4} (left) the dependence of the anomalous correlators with $\lambda$. One can see that the behaviour is linear with $\lambda$ in all the cases. The inset figures are given to show that the corresponding correlators do not vanish at $\lambda=0$. This is in fact true, as the non-vanishing values arise due to the $\kappa$ coupling, which leads to $\langle T_{0x}T_{0y}\rangle \sim \mu_5^3 \kappa$ and $\langle J_{0y}T_{0x}\rangle\sim \mu_5^2 \kappa$ at $\lambda=0$, with some contribution from $\Delta$. In the case of $\langle J_yJ_x\rangle$ the $\lambda$ dependence only arises in the massive case.
	
	Setting the values of $\Delta=\mu_5=0.1$, $\lambda=-1/(768\pi^2)$ and varying $\kappa$, we see a  similar kind of linear behaviour with $\kappa$. The effect of $\kappa$ is only seen in $\langle T_{0x}T_{0y}\rangle$, $\langle J_{0y}T_{0x}\rangle$ and $\langle J_yJ_x\rangle$ as shown in Fig.~\ref{figld4} (right) and Fig.~\ref{figkp1}. The non-anomalous correlators are independent of $\kappa$, and they are displayed in Fig.~\ref{figld1} of Appendix~\ref{App:B}. These correlators are in fact independent of both the parameters $\kappa$ and $\lambda$, and hence they are non-anomalous in nature even in the massive theory. This means that they do not contribute to anomalous transport, unlike the correlators studied above which are associated to anomalous conductivities. This can be seen in the Kubo formulae for anomalous conductivities, Eqs.~ (\ref{eqku1}) and (\ref{eqku2}), as these formulae involve Levi-Civita ($\epsilon_{ijz}$) symbols which runs over $i=j=\{x , y\}$. Hence, the correlators with $i=j=x$ and $i=j=y$ do not lead to anomalous transport effects.

\begin{figure}[t]
		\centering
		\begin{subfigure}{0.45\textwidth}
			\centering
			\includegraphics[width=1\linewidth]{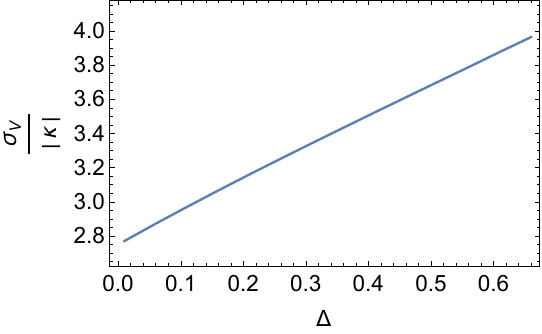}
		\end{subfigure}\hfill
		\begin{subfigure}{0.45\textwidth}
			\includegraphics[width=1\linewidth]{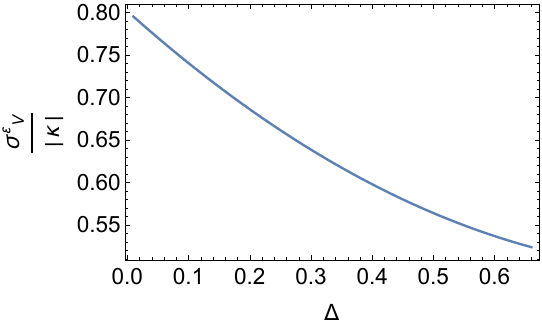}
		\end{subfigure}
		\centering
	\begin{subfigure}{0.45\textwidth}
		\centering
		\includegraphics[width=1\linewidth]{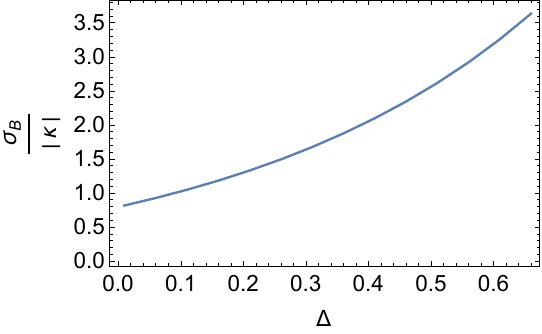}
	\end{subfigure}\hfill
	\begin{subfigure}{0.45\textwidth}
		\includegraphics[width=1\linewidth]{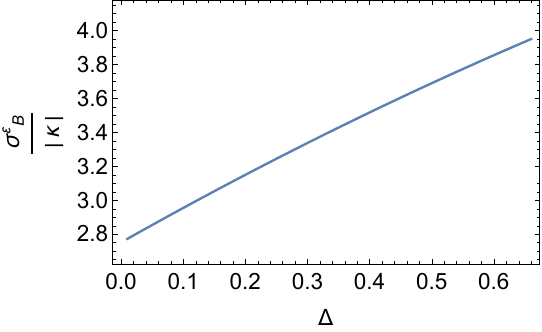}
	\end{subfigure}
		\caption{Upper panel: Plot of $\sigma_V$ (left) and $\sigma^\varepsilon_V $ (right) vs $\Delta$. Lower panel: Plot of $\sigma_B$ (left) and  $\sigma^\varepsilon_B $ (right) vs $\Delta$. We have considered $\mu_5= 0.15$ in all the panels.}
		\label{figsg1}
	\end{figure}

	\paragraph{Anomalous conductivities:} Finally, as a summary of the previous numerical results, we now present the anomalous conductivities which are computed with the Kubo formulas (\ref{eqku1}) and (\ref{eqku2}), i.e.
	\begin{align}
		\sigma_V &= -\lim_{k\to 0} \frac{1}{k} \textrm{Im} \langle J_x T_{0y}\rangle \,, \qquad \sigma_V^\varepsilon  = -\lim_{k\to 0} \frac{1}{k} \textrm{Im} \langle T_{0x} T_{0y}\rangle \,, \\
		\sigma_B &= - \lim_{k\to 0} \frac{1}{k} \textrm{Im} \langle J_x J_y \rangle \,, \qquad \sigma_B^\varepsilon = -\lim_{k\to 0}\frac{1}{k} \textrm{Im} \langle T_{0x} J_y \rangle \,.
	\end{align}
The results are displayed in Fig.~\ref{figsg1}. We can see from this figure that the chiral vortical conductivity and the chiral magnetic conductivity for energy current are the same either at zero or finite mass, i.e. $\sigma_V = \sigma^\varepsilon_B$, and these quantities increase with $\Delta$. We also see in this figure that the chiral vortical conductivity of energy current, $\sigma^\varepsilon_V$, decreases with $\Delta$ but the rate decreases rapidly. In the case of the chiral magnetic conductivity, $\sigma_B$, it increases with $\Delta$ as shown in Fig.~\ref{figsg1}.

Regarding the other dependences of the anomalous conductivities, for instance the dependence in the parameters $\kappa$ and $\lambda$, it would be sufficient to study them from Fig.~\ref{figld3} and Fig.~\ref{figld4}, as the two-point functions and the anomalous conductivities are related through Kubo formulae. We conclude that for a given value of $\mu_5$ and $\Delta$, the anomalous transport coefficients: $\sigma_V, \sigma^\varepsilon_B, \sigma_B$  and $\sigma^\varepsilon_B$; change linearly with the pure ($\kappa$) and mixed ($\lambda$) gauge-gravitational Chern-Simon couplings.  At the limit of vanishing mass, our results lead to $\frac{\sigma_B}{\mu_5 |\kappa|}\simeq 16/3$, which exactly coincides with the results in  \cite{Gynther:2010ed,Landsteiner:2011iq,Jimenez-Alba:2014iia} where $\alpha$ has been set to $\mu_5$ in both references~\footnote{$\alpha$ corresponds to the asymptotic value of the gauge field $A_t$ for $\rho \to 0$. In our case, we assume $\alpha = \mu_5$ for $\Delta = 0$.}. In order to reproduce the results of \cite{Gynther:2010ed} where they have set $\alpha = 0$, our $\kappa$ needs to be rescaled by a factor 3/2. Finally, let us emphasize that all the correlators involving the energy-momentum tensor are completely new results at finite mass $(\Delta \ne 0)$, i.e. $\sigma_V, \sigma_V^\varepsilon$ and $\sigma_B^\varepsilon$.
	
	\section{Discussion}
        \label{sec:Discussion}

	We have studied the anomalous and non-anomalous conductivities in the holographic St\"uckelberg model including both pure gauge and mixed gauge-gravitational anomaly terms. To access the sectors concerning the energy-momentum tensor we have to consider the full backreaction of the massive gauge field onto the metric tensor.  We have evaluated the numerical background solution and on this background, we have considered the fluctuations of the fields. From these fluctuations, we have calculated the different correlators and studied their behaviors with the relevant parameters of the model ($\mu_5$, $\Delta$, $\kappa$ and $\lambda$). 
	
	We have found that the correlators in the massless case match with previous results in the literature \cite{Amado:2011zx}. Later on, we have studied the dependence of these correlators with the mass of the gauge field, $m^2=\Delta(\Delta+2)$, and found that all the correlators explicitly depend on the mass for a given non-zero value of $\mu_5$. One of the results that it is important to emphasize here is that the non-anomalous correlators such as $\langle J_xJ_x\rangle$ and  $\langle T_{0x}J_x\rangle$ are non-vanishing in the massive theory for finite values of $\mu_5$. Moreover $\langle J_xJ_x\rangle$ is non-zero in this theory even for $\mu_5=0$, while $\langle T_{0x}J_x\rangle$ is vanishing for $\mu_5 = 0$ independently of the mass. These correlators are vanishing in the massless theory, independently of $\mu_5$. The mass of the gauge field highly enhances the absolute value of the correlators, and this gets translated into an enhancement of the anomalous conductivities. The behaviours of the correlators on the pure gauge and mixed gauge-gravitational Chern-Simon couplings, $\kappa$ and $\lambda$, were also studied. We found that the correlators  $\langle J_xJ_x\rangle,\langle J_xT_{0x}\rangle,\langle T_{0x}J_x\rangle$  and $\langle T_{0x}T_{0x}\rangle$ are independent of $\kappa$  and $\lambda$, and hence they are non-anomalous in nature. They do not contribute to the anomalous conductivities, as it can be seen from the Kubo formulae (\ref{eqku1}) and (\ref{eqku2}) as well. 
	
	Finally, we have computed the anomalous conductivities and studied their dependence with the mass of the gauge field ($m)$.  We have found that the chiral vortical conductivity, $\sigma_V$, and the chiral magnetic conductivity for energy current, $\sigma^\varepsilon_B$, are equal and increase with $\Delta$. One interesting result is that there are contributions to $\sigma_B$ coming from $\lambda$ in the massive theory, which was completely absent in the massless case. The conductivities $\sigma_B$, $\sigma_V$ and $\sigma^\varepsilon_B$  increase with $\Delta$, while the chiral vortical conductivity of energy current, $\sigma^\varepsilon_V$, decreases with~$\Delta$. We have explicitly checked that all our numerical results for the conductivities at finite mass tend to the known results at zero mass in the limit $\Delta \to 0$. For instance, it is known that at zero mass, the chiral magnetic conductivity is  $\sigma_B = -\frac{16}{3}\kappa \mu_5$ when $\alpha=\mu_5$, which implies that the ratio $-\frac{\sigma_B}{\kappa \mu_5} = 16/3$, independently of $\kappa$ and $\mu_5$. As one can see from Fig.~\ref{figsg1} (left), our numerics produces in this limit $\sigma_B/|\kappa| = 0.8$ for $\mu_5 = 0.15$, in agreement with the expected result. We have also checked the ratio $-\frac{\sigma_B}{\kappa \mu_5} = 16/3$ for other values of $\kappa$ and $\mu_5$.
	
	%One may note that in the limit of vanishing mass, the chiral magnetic conductivity tends to the value $\sigma_B = 8 \kappa \mu_5$, which exactly matches with our case of $\sigma_B$, which tends to $\sigma_B\sim0.4$ as mass tends to zero. With our value of $\mu_5=\frac{1}{4}$ and $\kappa=0.2$ this value of $\frac{\sigma_B}{\mu_5 \kappa}\simeq 8$ .  
	
	%This work can be extended in several ways. One possible extension could be considering the $U(1)_V \times U(1)_A$ set up with one of the gauge fields being non-anomalous, as it was done in Ref.~\cite{Megias:2016ery}. There the author has shown that there are anomalous contributions to the non-anomalous correlators as well, but the author has considered a probe limit, so the chiral vortex effects are not accessible. Initially one could study the massless case including the full backreaction on the metric, after that one might consider the massive case as well. It would be an interesting direction to pursue to see the interplay between the anomalous and non-anomalous terms in the set up of full backreacted background and these are the avenue that we would like to explore and address in our future work.
	
	This work can be extended in several ways. One possible extension could be to consider the $U(1)_V \times U(1)_A$ gauge group. There are some studies in holography with this gauge group, see e.g. Refs.~\cite{Jimenez-Alba:2014iia,Landsteiner:2013aba,Megias:2016ery}. However, in these works: i) either the probe limit has been considered so that the chiral vortical effect and the transport conductivities in the energy-momentum tensor are not accessible, or ii) they correspond to studies for massless gauge bosons. In particular, it would be interesting to study the interplay between the anomalous and non-anomalous currents in the set-up of the full backreacted background of Ref.~\cite{Megias:2016ery}, both for massless and massive gauge bosons. We will explore these and other issues in future works.

	%set up with one of the gauge fields being non-anomalous, as it was done in Ref.~\cite{Megias:2016ery}. There the author has shown that there are anomalous contributions to the non-anomalous correlators as well, but the author has considered a probe limit, so the chiral vortex effects are not accessible. Initially one could study the massless case including the full backreaction on the metric, after that one might consider the massive case as well. It would be an interesting direction to pursue to see the interplay between the anomalous and non-anomalous terms in the set up of full backreacted background and these are the avenue that we would like to explore and address in our future work.
	
\section*{Acknowledgments}
	We would like to thank Karl Landsteiner for enlightening discussions. E.M. is grateful to Manuel Valle for collaboration in the early stages of this work. N.R. thanks the Instituto de F\'{\i}sica Te\'orica UAM/CSIC, Spain, for its hospitality and partial support during his research visits in the final stages of this work. The works of N.R. and E.M. are supported by the project PID2020-114767GB-I00 funded by MCIN/AEI/10.13039/501100011033, by the FEDER/Junta de Andaluc\'{\i}a-Consejer\'{\i}a de Econom\'{\i}a y Conocimiento 2014-2020 Operational Program under Grant A-FQM-178-UGR18, and by the Ram\'on y Cajal Program of the Spanish MCIN under Grant RYC-2016-20678. The work of E.M. is also supported by Junta de Andaluc\'{\i}a under Grant FQM-225.

%\pagebreak
%\newpage

\begin{appendices}

\

\section{Explicit expressions for the functions $\Omega(\rho)$ and $\Phi_j(\rho)$}
\label{App:A}

These functions have been introduced in the equations of motion of the fluctuations (\ref{fleq1})-(\ref{fleq2}). Their explicit expressions are given by
\begin{eqnarray}
\begin{split}
&\Omega(\rho)=\frac{4 \Big[g_{\tau \tau }(\rho ) \left(g_{xx}^\prime(\rho )+2 \rho  g_{xx}^{\prime\prime}(\rho )\right)+\rho  g_{xx}^\prime(\rho ) g_{\tau \tau }^\prime(\rho
					)\Big]}{\sqrt{g_{xx}(\rho )} g_{\tau \tau }(\rho ){}^{3/2}} - 8\rho\frac{g_{xx}^\prime(\rho )^2}{g_{xx}(\rho)^{3/2}} \sqrt{g_{\tau \tau }(\rho )} \\
&+\frac{\sqrt{g_{xx}(\rho )}}{g_{\tau\tau }(\rho ){}^{5/2}}\left[ 4 \rho  g_{\tau \tau }^\prime(\rho ){}^2-4 g_{\tau \tau }(\rho) \left(g_{\tau \tau }^\prime(\rho )+2 \rho  g_{\tau \tau }^{\prime\prime}(\rho )\right)\right]  \,,
\end{split}
\end{eqnarray}
		and
\begin{eqnarray}
\begin{split}
& \Phi_{j}(\rho)=b_j^\prime(\rho ) \Bigg[-\frac{8  \rho ^2 \sqrt{g_{\tau \tau }(\rho )}
					g_{xx}^\prime(\rho ){}^2}{g_{xx }(\rho ){}^{7/2}}+\frac{4  \rho  \Big(g_{\tau
						\tau }(\rho ) \Big(g_{xx }'(\rho )+2 \rho  g_{xx }''(\rho )\Big)+\rho 
					g_{xx }^\prime(\rho ) g_{\tau \tau }'(\rho )\Big)}{g_{xx}(\rho ){}^{5/2}\sqrt{g_{\tau \tau }(\rho )}}\\
&-\frac{4  \rho  \Big(g_{\tau \tau }(\rho )\Big(g_{\tau \tau }'(\rho )+2 \rho  g_{\tau \tau }''(\rho )\Big)-\rho  g_{\tau \tau}'(\rho ){}^2\Big)}{g_{xx }(\rho ){}^{3/2} g_{\tau \tau }(\rho
					){}^{3/2}}\Bigg] \\
&+b_j(\rho ) \Bigg[\frac{8 \rho ^2 \sqrt{g_{\tau \tau }(\rho)} g_{xx}^\prime(\rho ){}^3}{g_{xx }(\rho ){}^{9/2}} -\frac{8  \rho \sqrt{g_{\tau \tau }(\rho )} g_{xx }^\prime(\rho )}{g_{xx }(\rho ){}^{7/2}} \Big(g_{xx }'(\rho )+2 \rho g_{xx }''(\rho )\Big) \\
&+\frac{4 \rho  \Big(\rho g_{xx }''(\rho ) g_{\tau \tau }'(\rho )-\rho  g_{xx }^\prime(\rho ) g_{\tau \tau
					}''(\rho )+3 g_{\tau \tau }(\rho ) g_{xx }''(\rho )+2 \rho  g_{xx}{}^{(3)}(\rho ) g_{\tau \tau }(\rho )\Big)}{g_{xx }(\rho ){}^{5/2}
					\sqrt{g_{\tau \tau }(\rho )}}\\
&-\frac{4  \rho  \Big(-2 g_{\tau \tau }(\rho ) g_{\tau
						\tau }'(\rho ) \Big(g_{\tau \tau }'(\rho )+2 \rho  g_{\tau \tau }''(\rho )\Big)+2
					\rho  g_{\tau \tau }'(\rho ){}^3+g_{\tau \tau }(\rho ){}^2 \Big(3 g_{\tau \tau
					}''(\rho )+2 \rho  g_{\tau \tau }{}^{(3)}(\rho )\Big)\Big)}{g_{xx }(\rho
      ){}^{3/2} g_{\tau \tau }(\rho ){}^{5/2}}\Bigg]
\end{split}
\end{eqnarray}

\begin{eqnarray}
  \begin{split}
&+ h^j{}_t^\prime(\rho ) \Bigg[ B_t^\prime(\rho ) \left(\frac{16  \rho ^2 g_{xx }^\prime(\rho
      )}{g_{xx }(\rho ){}^{3/2} \sqrt{g_{\tau \tau }(\rho )}}+\frac{8\rho \Big(g_{\tau \tau }(\rho )-\rho  g_{\tau \tau }^\prime(\rho )\Big)}{\sqrt{g_{xx}(\rho )} g_{\tau \tau }(\rho ){}^{3/2}} \right) + \frac{8  \rho ^2 B_t^{\prime\prime}(\rho)}{\sqrt{g_{xx }(\rho )} \sqrt{g_{\tau \tau }(\rho )}}\Bigg]  \nonumber \\
&+h^j{}_t^{\prime\prime}(\rho) \frac{8\rho^2 B_t'(\rho )}{\sqrt{g_{xx }(\rho )} \sqrt{g_{\tau \tau }(\rho)}}  \,. \nonumber
			\end{split}
		\end{eqnarray}
		%        \pagebreak

\

\section{Some additional results for the non-anomalous correlators}
\label{App:B}

\begin{figure}[h]
			\centering
			\begin{subfigure}{0.45\textwidth}
				\centering
				\includegraphics[width=1\linewidth]{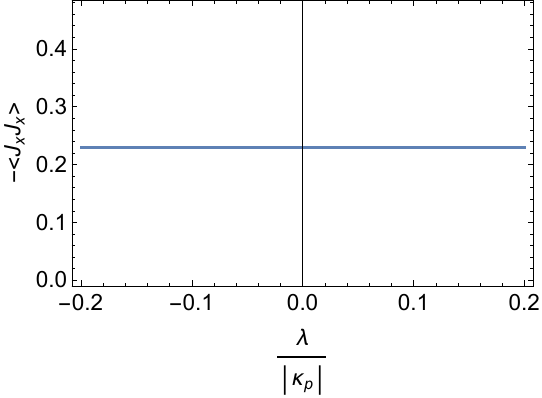}
			\end{subfigure}\hfill
			\begin{subfigure}{0.45\textwidth}
				\includegraphics[width=1\linewidth]{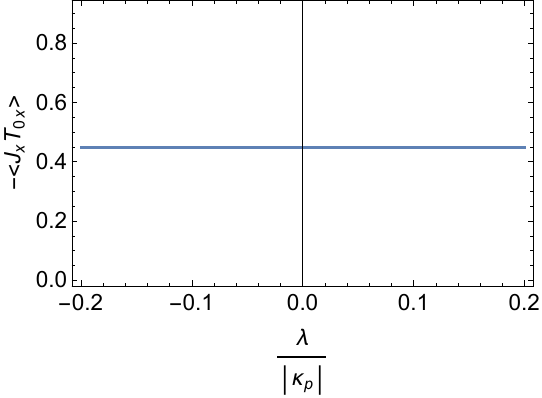}
			\end{subfigure}
			\centering
		\begin{subfigure}{0.45\textwidth}
			\centering
			\includegraphics[width=1\linewidth]{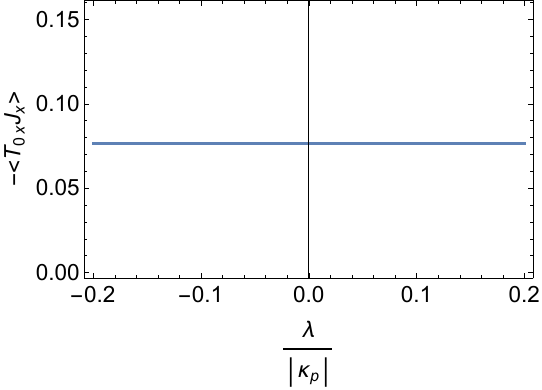}
		\end{subfigure}\hfill
		\begin{subfigure}{0.45\textwidth}
			\includegraphics[width=1\linewidth]{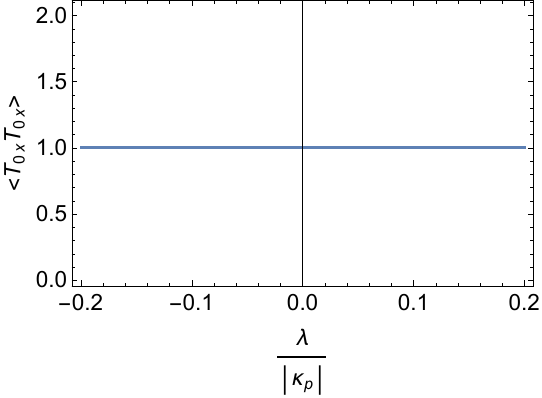}
		\end{subfigure}
			\caption{Upper panel: plot of the correlator $\langle J_{x}J_{x}\rangle $ (left) and  $\langle J_{x}T_{0x}\rangle $ (right) vs $\lambda/|\kappa_p|$. Lower panel: plot of the correlator $\langle T_{0x}J_{x}\rangle $ (left) and $\langle T_{0x}T_{0x}\rangle $ (right) vs $\lambda/|\kappa_p|$. We have considered $\mu_5=0.1$, $\Delta=0.1$ and $\kappa_p =-1/(32\pi^2)$ in all the panels.}
			\label{figld2}
		\end{figure}

\begin{figure}[t]
		\centering
		\begin{subfigure}{0.45\textwidth}
			\centering
			\includegraphics[width=1\linewidth]{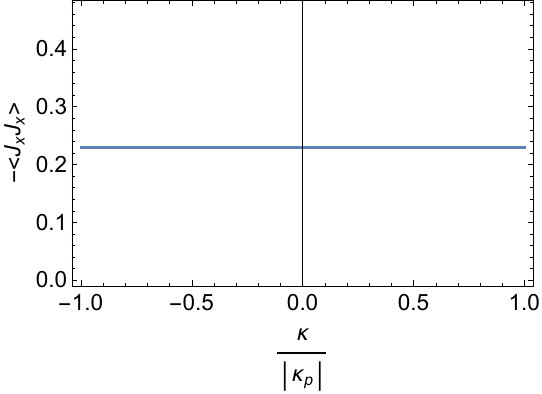}
		\end{subfigure}\hfill
		\begin{subfigure}{0.45\textwidth}
			\includegraphics[width=1\linewidth]{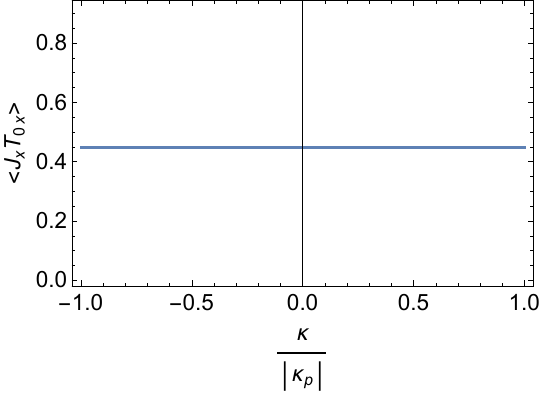}
		\end{subfigure}
	\centering
	\begin{subfigure}{0.45\textwidth}
		\centering
		\includegraphics[width=1\linewidth]{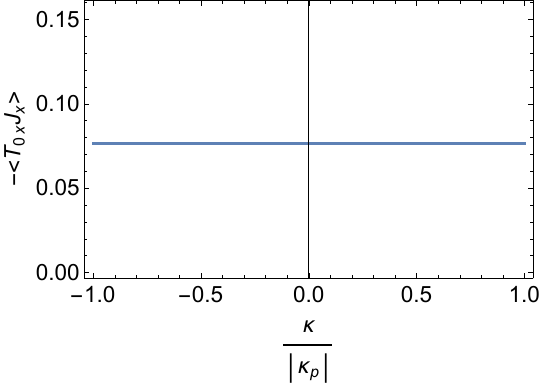}
	\end{subfigure}\hfill
	\begin{subfigure}{0.45\textwidth}
		\includegraphics[width=1\linewidth]{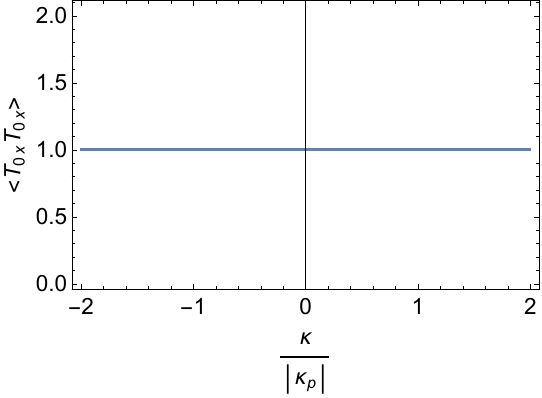}
	\end{subfigure}
		\caption{Upper panel: plot of the correlator $\langle J_{x}J_{x}\rangle$ (left) and $\langle J_{x}T_{0x}\rangle $ (right) vs $\kappa/|\kappa_p|$. Lower panel: plot of the correlator $\langle T_{0x}J_{x}\rangle $ (left) and $\langle T_{0x}T_{0x}\rangle $ (right) vs $\kappa/|\kappa_p|$. We have considered $\mu_5=0.1$, $\Delta=0.1$, $|\kappa_p|=1/(32\pi^2)$ and $\lambda = -1/(768 \pi^2)$ in all the panels.}
		\label{figld1}
	        \end{figure}

We show in this Appendix the numerical results for the non-anomalous correlators as a function of the anomalous parameters $\kappa$ and $\lambda$, in the massive case $(\Delta \ne 0)$. The correlators $\langle J_x J_x \rangle$, $\langle J_x T_{0x} \rangle$, $\langle T_{0x} J_x \rangle$ and $\langle T_{0x} T_{0x}\rangle$, are displayed in Figs.~\ref{figld1} and \ref{figld2}. These correlators turn out to be constant in both $\kappa$ and $\lambda$. The lack of dependence in these parameters implies that they lead to non-anomalous transport effects.

	\end{appendices}

\end{document}